# Publication and citation statistics for UK astronomers

Alexander Blustin

UCL Mullard Space Science Laboratory

**Abstract**

This article presents a survey of publication and citation statistics for 835 UK professional astronomers: the majority of academics and contract researchers within the UK astronomical community. I provide histograms of these bibliometrics for the whole sample as well as of the median values for the individual departments. I discuss the distribution of top bibliometric performers in the sample, and make some remarks on the usage of bibliometrics in a real-world assessment exercise.

**Introduction**

There has been much interest recently in the use of bibliometric measures to assess the quality and quantity of research in UK universities (Roberts 2003; HM Treasury 2006). Bibliometrics – measures of the numbers of publications and citations – are an unambiguous numerical measure of research output, free from the potential subjectiveness of grant panels and much less time-intensive. Many have questioned, though, whether bibliometrics provide a useful measure of quality (e.g. Pijpers 2006).

Either way, it is likely that future research assessment exercises will make much greater use of publication and citation data. In the context of UK astrophysical research, it is interesting to examine the national distribution of these quantities and what this implies about how bibliometrics should and should not be used. In this article, I present the results of a survey of the publication and citation statistics of 835 astronomers in UK universities – the majority of the UK professional astronomical community – and draw some conclusions about the usage of bibliometrics in practice.

**Data collection**

I compiled lists of academic staff and contract researchers from the websites of all UK astronomy groups with at least five academic staff – 30 groups in total. Where there exist multiple large astronomy-related departments within the same university, with significantly different areas of activity, I have treated them separately. This is the case for the three departments at Cambridge and two at UCL which carry out astronomical research. I have taken "academic staff" to include those holding senior research fellowships (e.g. Royal Society Research Professors and University Research Fellows,

PPARC/STFC Advanced Fellows, Academic Fellows and equivalent), and also emeritus staff who have published in refereed journals within the qualifying time interval. The contract researchers include those employed in project and support roles, who often have a significant research output, but not computing staff, programmers or engineers. All students, visiting and honorary staff were excluded. I include scientists working in all branches of astronomy-related research, including cosmology and gravitation, astro-particle physics and theoretical astrophysics. Space plasma physicists, planetary geologists and instrumentation-only groups are omitted, as are solar physicists except where they have an interest in stellar phenomena. The lists contain a total of 440 academics and 395 contract researchers.

I used the NASA Astrophysics Data Service (ADS) astronomy and physics archives to collect publication and citation numbers for each academic and contract researcher listed, using papers published between April 2001 and April 2006. The census epoch for staff lists and citation counts is December 2006. I collected data for both total publications and first author publications. Only refereed research papers in mainstream technical journals were included; book reviews, obituaries, journal editorials and articles published in A&G and Observatory were excluded. For each individual I took great care to separate out the publications of scientists with the same name, and to take account of name changes and cases where different variants of first names and initials were used.

While great care has been taken in obtaining the data presented here, there are several possible sources of error in an exercise of this kind. The biggest pitfall is name variations, different scientists with the same name, and inconsistency in use of middle names and initials. I dealt exhaustively with the many cases I encountered, but it is possible that I might have missed some instances. I did not exclude self-citations, since the name-confusion issue makes it very hard to tell whether this is being done correctly. Another source of uncertainty is the web pages from which I obtained staff lists, some of which were out of date, inaccurate or vague. Some did not state if the people listed were academic staff or contract researchers, so I worked this out from secondary sources. It is possible that I have wrongly included instrumentation-only staff in some institutes where they are not marked out as belonging to a separate group. The entire dataset relies on the accuracy and comprehensiveness of ADS; the principal concern here would be that the publication and citation numbers could be underestimated for mathematical areas that are perhaps not fully covered by the database. Another unusual feature of this survey is the inclusion of all academic staff listed as members of the research groups; in any real assessment exercise, a certain proportion of them would be classified as "teaching-only" and hence excluded from consideration. I have no way of distinguishing these staff here, and this division is a contentious issue anyway.

**National performance**

In figure 1 I plot the histograms of the cumulative percentages of scientists achieving more than a given number of publications or citations. For each bibliometric, I also mark the levels corresponding to 50% and 95% of scientists, and list the relevant numbers in table 1. Academics and contract researchers are plotted separately.

There is a wide range in publication output. At the top end, one individual had 178 publications, while there were 11 academics (2.5% of the total) with no publications at all over five years. These, presumably, are genuinely teaching-only or administration-only staff. In terms of first author publications, the top performer wrote an average of seven per year – while 89 academics (20%) had not produced any first author papers in five years.

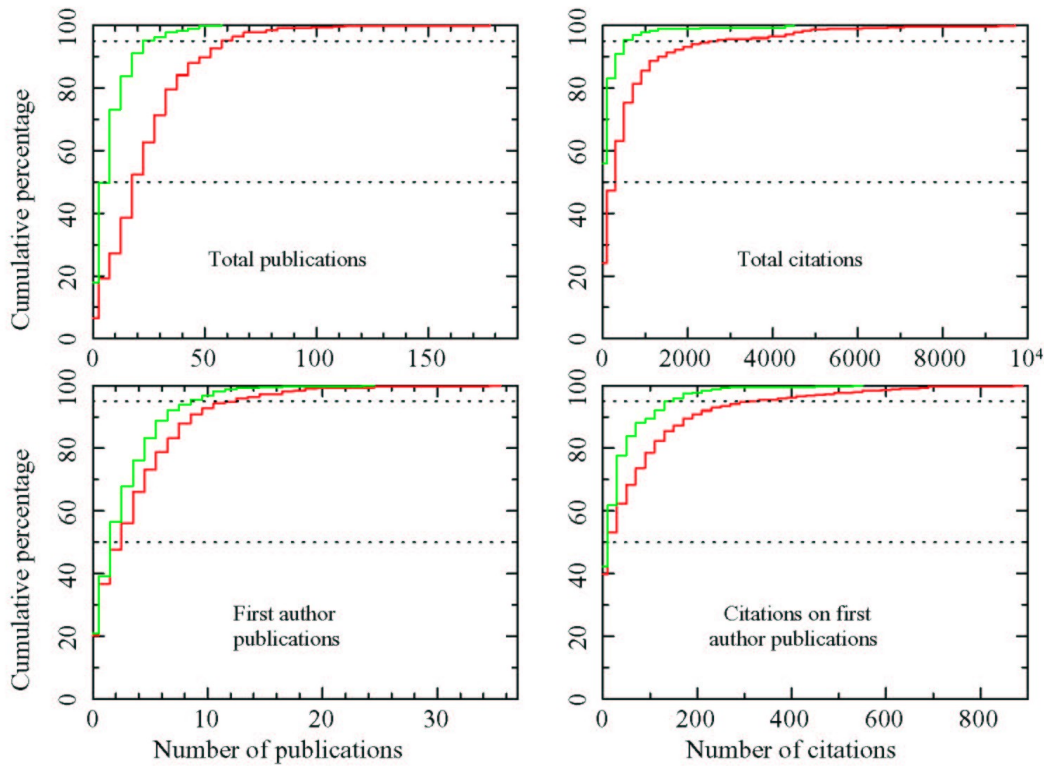

*Figure 1: Cumulative percentages of academics (red) and contract researchers (green) achieving less than or equal to a given bibliometric. The dotted lines represent 50% and 95% of staff.*

The greatest numbers of publications and first author publications by contract researchers were 58 and 24 respectively; 26 contract researchers (6.6%) had no publications and 83 (21%) no first author publications in five years.

Table 1 lists the numbers of publications and citations corresponding to the 95% and 50% levels, obtained from ordered lists of all the scientists in the sample. The median academic in UK astronomy (i.e. corresponding to the 50% level) produces, on average, 4.4 papers per year of which they are the first author of 0.6. The median contract researcher has 1.6 papers per year, 0.4 of which are first author. The difference between total publication rates is presumably due to the more established collaboration networks of the academics. The rates of first author publications are surprisingly similar; either experience does not count for much, or its advantages are offset by the demands of teaching and administration.

| **Academics** | **Top 50%** | **Top 5%** |
|---|---|---|
| Total publications | $\geq 22$ | $\geq 62$ |
| Total citations | $\geq 323$ | $\geq 2734$ |
| First author publications | $\geq 3$ | $\geq 13$ |
| First author citations | $\geq 25$ | $\geq 337$ |
| **Contract researchers** | **Top 50%** | **Top 5%** |
| Total publications | $\geq 8$ | $\geq 27$ |
| Total citations | $\geq 80$ | $\geq 681$ |
| First author publications | $\geq 2$ | $\geq 9$ |
| First author citations | $\geq 16$ | $\geq 152$ |

*Table 1: Identifying the high performers. Bibliometric performance better than or equal to the numbers in this table indicates that a scientist is in the top 50% or top 5% of academics and contract researchers in the sample.*

**Comparing departments**

Although, nationally, there is a wide range of bibliometric performance, one can ask whether this is effectively averaged out within individual departments, or whether there are systematic differences between the bibliometric performance of different departments. The simplest comparison is between the performance of the "median staff members" in each department. This is shown in figure 2, where I plot histograms of the median bibliometric values for the academics in all of the groups in my sample.

The median bibliometric activity of different groups is indeed very variable. It is interesting that the distribution of total citations is bimodal, whereas the total publications plot has a single peak, which may illustrate that, however much effort goes into paper production, different sub-fields within astronomy have completely different citation patterns that may depend, for example, upon global trends reaching far beyond the UK.

The situation with first author papers is more complicated, which is unsurprising because the concept of "first authorship" implies entirely different things in different sub-fields of astronomy. It is a meaningless measure for departments engaged in, for example, dark matter experiments where long, alphabetized author lists are the norm. Conversely, it may be a useful indicator of research effort in a field where the first author is the person who has written the paper and done most of the work. In the latter case, citations on first author papers (taking into account the citation pattern of the sub-field) will be a genuine measure of the influence of the work of an individual scientist.

Another source of bias in these median bibliometrics is the presence of staff who are primarily teachers or administrators, with little involvement in research. About 15% of academics published one or fewer papers per year on average. Of these, 64% belonged to departments with 20 or more academics, although these large departments only accounted for 50% of the total academic population. 17% of low-publication-rate academics were in departments with 10 or fewer academic staff; these departments contained 22% of the national total.

Relatively research-inactive staff are therefore more likely to be found in large departments, which could be a reflection of large student intakes, or simply that the culture of a large department allows more scope for specialization in the different aspects of academic life. It is also possible that, in smaller groups, individual staff are far more visible and under greater pressure to sustain a productive research programme.

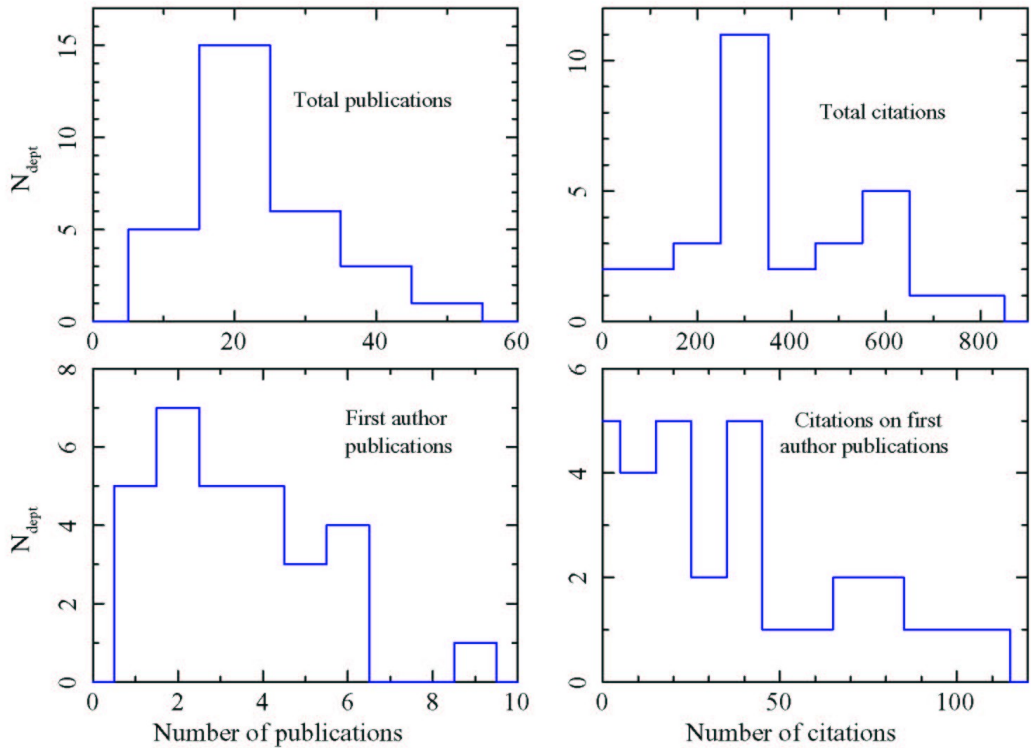

*Figure 2: Histograms of the median bibliometrics for each department in the sample. The y-axis is the number of departments with a median bibliometric in the given range.*

Median bibliometrics provide no recognition for a small world-class research group within a larger teaching-intensive department. Plotting the bibliometrics for each member of staff in the six largest groups in my sample (figure 3) illustrates this well. Many of the departments contain an "inner core" of very high bibliometric productivity, which often, though not always, signals the presence of cosmologists; these plots also give an indication of the existence of co-authorship groups.

The location of centres of excellence presumably has some relation to the presence of top bibliometric performers, if bibliometrics are a measure of quality. I have identified the top 5% of scientists for each bibliometric, again treating academics and contract researchers separately, and looked at how they are distributed between the departments in my sample. The results of this exercise are summarized in table 2.

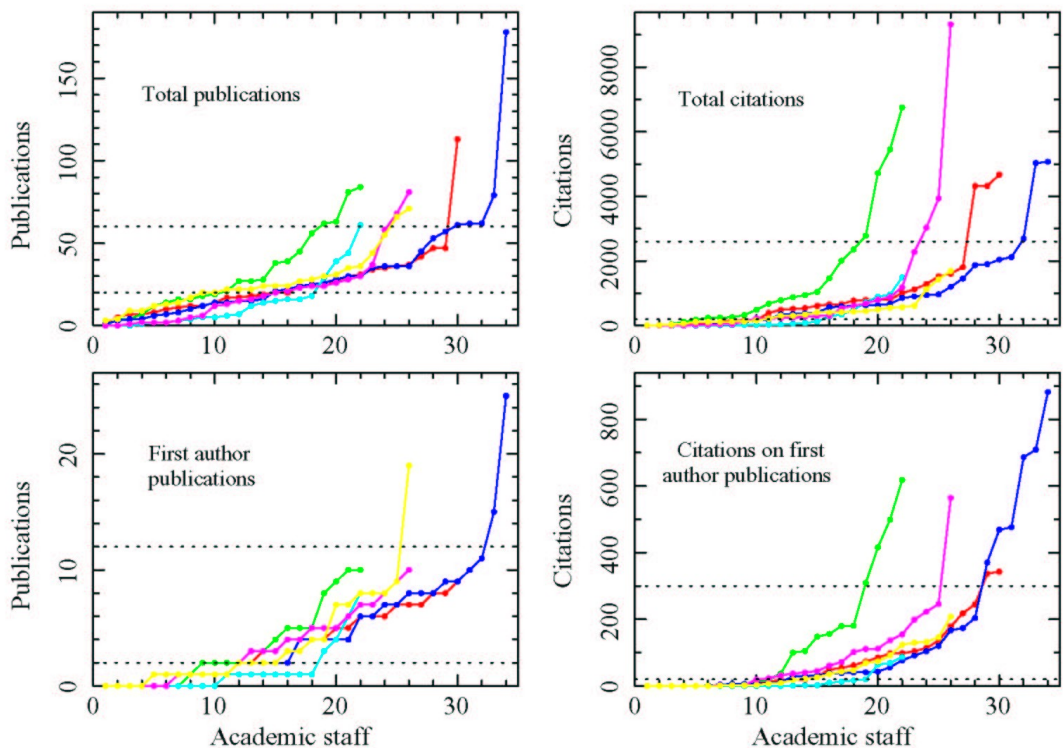

*Figure 3: Plots of each bibliometric for the members of academic staff in the six largest departments in the sample. Each colour represents a different department; in each case, the relevant bibliometric in order of increasing score within the department is plotted. The order of individual staff members usually, therefore, varies between the plots. The dotted lines represent the bibliometric scores for the top 5% and top 50% of academics in the whole sample.*

Overall, 66 academics and 55 contract researchers achieved in the top 5% of their respective distributions on at least one bibliometric. These top performers are, evidently, widespread: 26 out of 30 departments (24 out of 27 universities) are represented among this group. The distribution is, however, highly peaked: 19 of these scientists are accounted for by a single department, which also happens to be the largest department in my sample.

We can become more selective by defining the best bibliometric performers to be those achieving in the top 5% on at least two measures, which narrows the field significantly. For academics, the 19 people in this category are based in 11 out of 30 departments, or 10 out of 27 universities. At the very top, only three departments host academics scoring in the top 5% on three or four bibliometrics. The single academic who appears in the top 5% for all measures is based in the department containing the largest number of top scorers overall.

A. J. Blustin 7

| Academics | No. of staff | No. (%) of the 30 departments that contain these staff | No. (%) of the 27 universities that contain these staff |
|---|---|---|---|
| Top 5% for ≥ 4 bibliometrics | 1 | 1 (3.3%) | 1 (3.7%) |
| Top 5% for ≥ 3 bibliometrics | 5 | 3 (10%) | 3 (11%) |
| Top 5% for ≥ 2 bibliometrics | 19 | 11 (37%) | 10 (37%) |
| Top 5% for ≥ 1 bibliometric | 66 | 23 (77%) | 22 (81%) |
| **Contract researchers** | No. of staff | No. (%) of the 30 departments that contain these staff | No. (%) of the 27 universities that contain these staff |
| Top 5% for ≥ 4 bibliometrics | 0 | 0 (0%) | 0 (0%) |
| Top 5% for ≥ 3 bibliometrics | 9 | 7 (23%) | 7 (26%) |
| Top 5% for ≥ 2 bibliometrics | 22 | 12 (40%) | 12 (44%) |
| Top 5% for ≥ 1 bibliometric | 55 | 20 (67%) | 19 (70%) |

*Table 2: Where are the high performers? The distribution of bibliometric top performers among the astronomy groups in the sample; departments and universities are treated separately, since two institutions contain more than one large astronomy-related department with significantly different areas of activity (see section on data collection). The second column gives the cumulative numbers of staff scoring in the top 5% on the number of bibliometrics listed in column one. Columns two and three give the number (and percentage) of departments and universities respectively across which these staff are distributed.*

**Conclusions**

I have presented publication and citation bibliometrics for a sample of 440 academics and 395 contract researchers from 30 research groups at 27 UK universities; this study covers the vast majority of the UK professional astrophysics community. Having gone through the process of collecting this information and discussing it with various interested parties, I would like to present a few opinions about how bibliometrics should or should not be used in judging the quality of astrophysical research.

The biggest problem in collecting the data using ADS is that of name confusion; different scientists having the same surname and even initials, and inconsistent usage of initials and middle names. The ideal solution would be to give every astrophysicist a unique bibliometric identifier. In the absence of this, I believe that these data should be collected by scientists within the individual departments, who know the people concerned, the names they use and the research areas they work in, and not by a central bureaucrat who would have extreme difficulty in disentangling the confusion. Evidently, this also avoids the margin of error in this survey that is due to compiling staff lists from potentially inaccurate departmental websites.

There should be an unambiguous policy on what types of publications are included in the exercise, and preferably an explicit list of which journals should be included and excluded. I have limited this study to research papers published in mainstream refereed technical journals; different citation databases may have different definitions as to which journals meet this description.

There also need to be clear guidelines as to who is included in any real bibliometrics exercise, since there are many marginal cases, and thus opportunities for error and games-playing. The rules should cover staff employed on hardware projects who might not, technically, be paid to do research; also research students, "teaching" staff who publish research, visiting staff, honorary staff, and emeritus staff.

Another difficult case is that of the college fellows at the Oxbridge universities, who have widely varying levels of involvement in research groups. In this study I find that there are very few astronomy academics, nationally, who are genuinely teaching-only in the sense of never producing published research. Instead of conveniently excluding low-performing staff under the teacher designation, I would favour a quality profiles approach which accepts that a certain proportion of staff are likely to spend more time on teaching and administration, without ignoring the contribution that they make to the research of a department. In an ideal world, research output would be just one of the ways in which an academic would be recognized for the purposes of funding and promotion.

There has to be a policy on whether the exercise can include publications and citations earned abroad, or at a different institute. If the aim of the exercise is to assess the quality of individuals employed by a department at a given point in time, then it is reasonable to include these publications. If, on the other hand, the ultimate goal is to assess the quality of work done in the particular research environment of a given department, then science published whilst working elsewhere would have to be excluded. This latter case could have interesting implications for the academic job market.

As stated above, the concept of first authorship is not relevant to some astrophysical disciplines but very important to others. The bibliometrics chosen to judge research output in a field should be appropriate for that field. Different research areas also have widely different citation patterns, which are closely related to the current fashions in international astrophysics; citations will be an indicator of quality so long as the fashionable research areas are those that are novel, important and done well. In any case, it is important that highly popular fields do not drive out funding for emerging or niche areas that may, perhaps, become the high-profile science of the future. Peer review may be the only mechanism for keeping this balance.

Can bibliometrics be applied to instrumentation groups? What adjustments are necessary for astronomers who, although they publish science papers, spend much of their time building instruments and satellite missions? I would suggest a system whereby a department received a bibliometric royalty on the hardware it built, according to the volume of publications and citations in the international literature based on data from the hardware. This would ensure that hardware groups were rewarded for building instruments to do timely, high profile science. It would also benefit the broader community by providing a powerful incentive for rapid, high-quality calibration, and energetic promotion of the instruments to ensure that they were widely used.

*Alexander Blustin is an STFC Postdoctoral Fellow at UCL Mullard Space Science Laboratory.*

*Acknowledgements: I wish to acknowledge the many interesting discussions that I have had with a range of scientists on these issues, and am most grateful for all of their thoughtful comments and suggestions, particularly those relating to real-world instances of the use and misuse of bibliometric data. This survey has made use of the NASA Astrophysics Data System (ADS) bibliographic database. I acknowledge the support of an STFC Postdoctoral Fellowship.*

*Disclaimer: Due to the uncertainties and unusual features of this survey as stated above, it is not intended to be used as a basis for any real assessment exercise. All views expressed in this article are my own, and do not reflect the corporate position of UCL, MSSL or STFC.*